\documentclass[eqsecnum,floats,aps,prd,floatfix,titlepage,tightenlines]{revtex4} 

\usepackage{graphicx}
\usepackage{graphics}
\usepackage{bm}
\usepackage{amssymb}
\usepackage{amsmath}
\usepackage{mathrsfs}

\begin{document}
\title{ Numerical Simulation of Quantum Field Fluctuations} \author{Emily R. Taylor $^{1\dagger}$} \author{Samuel Yencho  $^{2*}$}
 \author{L. H. Ford$^{1\ddagger}$} 
\affiliation{$^1$Institute of Cosmology, Department of Physics and Astronomy, Tufts University, Medford, Massachusetts 02155, USA\\
$^2$Department of Physics, Applied Physics,  and Astronomy,  Rensselaer Polytechnic Institute, Troy, New York 12180, USA}

\begin{abstract}
The quantum fluctuations of fields can exhibit subtle correlations in space and time. As the interval between a pair of measurements varies, the
correlation function can change sign, signaling a shift between correlation and anti-correlation. A numerical simulation of the fluctuations requires
a knowledge of both the probability distribution and the correlation function. Although there are widely used methods to generate a sequence of random
numbers which obey a given probability distribution, the imposition of a given correlation function can be more difficult. Here we propose a simple method
in which the outcome of a given measurement determines a shift in the peak of the  probability distribution, to be used for the next measurement. We illustrate
this method for three examples of quantum field correlation functions, and show that the resulting simulated function agree well with the original, analytically
derived function. We then discuss the application of this method to numerical studies of the effects of correlations on the random walks of test particles
coupled to the fluctuating field.
\end{abstract}

\maketitle

\baselineskip=14pt

\section{Introduction}
\label{sec:intro}

Although vacuum fluctuations of a quantized field, such as the electromagnetic field, are formally infinite, careful treatments of the contributions
of high frequency modes lead to finite observable effects. Two examples are the Casimir effect and the Lamb shift. Welton~\cite{Welton48} has 
given a simple argument which illustrates that the dominant contribution to the Lamb shift is due to the effects of Brownian motion of an electron
responding to vacuum electric field fluctuations. In recent years, there has been interest in various physical effects which might be produced by
the quantum fluctuations of linear operators, such as the electric field, or of quadratic operators, such as the energy density. For a recent review,
see Ref.~\cite{LF22}.

Although quantum field fluctuations are usually treated with analytic methods, such as the calculation of variances, numerical simulations can also 
play a role. This was done, for example, by Carlip {\it et al}~\cite{Carlip11,Carlip20} to study the effects of quantum stress tensor fluctuations upon
the focussing of light rays. Here a model in two-dimensional spacetime was used, with the fluctuations satisfying a non-Gaussian probability
distribution given in Ref.~\cite{FFR10}.
However, as  Carlip {\it et al} note, their simulation ignores the possible correlations between fluctuations in different spacetime regions.
More generally, it is desirable to find methods to include correlations in simulations, as quantum fields can exhibit subtle correlations and anti-correlations. 
These are discussed, for example, in Refs~\cite{Yf04,PF11}.

There are statistical methods for introducing correlations into autoregression analyses. An example is the Monte Carlo method discussed in 
Ref.~\cite{CTY94}. However, we are not aware of any use of these methods in the literature to study quantum field fluctuations. The purpose of this paper 
is to discuss the correlation functions for linear field operators and to propose a method for introducing these correlations into a numerical
simulation. The outline of this paper is as follows: Sect.~\ref{sec:corr} will introduce and discuss three examples of quantum field correlation functions.
Our proposed method for implementing correlations in a simulation will be introduced in Sect.~\ref{sec:Sims}. The possible application of the method 
to the study of the effects of correlations on Brownian motion will be considered in Sect.~\ref{sec:Brownian}. Our results will be summarized and discussed
in Sect.~\ref{sec:final}.

\section{Correlation Functions}
\label{sec:corr}

In this section, we will discuss correlation functions for quantum fields, and give three explicit examples of physical interest. In general, a correlation function may be defined as
\begin{equation}
 C(t,{\bf x}; t', {\bf x'};) = \langle   \psi(t,{\bf x})\, \psi(t',{\bf x'})  \rangle   -  \langle   \psi(t,{\bf x}) \rangle  \,\langle    \psi(t',{\bf x'})   \rangle\,.
 \label{eq:corr-def}
 \end{equation}
where $  \psi(t,{\bf x})$ is a field operator at spacetime point $ (t,{\bf x})$, and $\langle \; \rangle$ denotes an expectation value in a selected quantum state.
This function describes the correlations in the fluctuations of $\psi$ between different spacetime regions.

In this paper, we restrict our attention to field operators at a fixed point  ${\bf x}$ in space, and drop explicit mention  of this point. We also assume that the mean field 
value vanishes, so
 \begin{equation}
   \langle   \psi(t,{\bf x}) \rangle =   \langle   \psi(t) \rangle = 0\, .
 \end{equation}
We also consider cases where the correlation function depends only upon the time difference, $t - t'$, and write
 \begin{equation}
 C = C(t-t') =  \langle   \psi(t)\, \psi(t')  \rangle \, .
  \label{eq:corr}
 \end{equation}

\subsection{Massless Scalar Field}
\label{sec:scalar}

In this subsection, we treat the vacuum fluctuations of a massless scalar field in four-dimensional Minkowski spacetime, where
 \begin{equation}
 C_0(t-t') = -\frac{1}{4\pi^2(t-t')^2} \,. 
 \label{eq:C0}
 \end{equation}
The $t=t'$ limit, $C_0(0)$, is singular due to the short distance divergences of a quantum field. Here we adopt the viewpoint that measurements of 
vacuum fluctuations at a single spacetime 
point are unphysical, and all measurements require an average over a finite spacetime region, or at least a finite time or space interval. 
For this purpose, define a time averaged
field operator by
\begin{equation}
 \bar\psi(t_0)=\int_{-\infty}^{\infty}dt \, \psi(t) \, f(t-t_0)\,,
 \end{equation}
where $f(t)$ is a sampling function peaked at $t =t_0$ which is normalized by  $\int_{-\infty}^{\infty}dt \, f(t) = 1$. As a convenient example, we use a 
Lorentzian form
 \begin{equation}
 f(t)=\frac{\tau}{\pi\left(t^2+\tau^2\right)}\,,
 \label{eq:Lorentz}
 \end{equation}
 where $\tau$ is the characteristic duration of the time averaging. The effect of the averaging will be to suppress the contributions of high frequency modes,
 those whose periods are short compared to $\tau$.

Let 
\begin{equation}
C(t_0)=\langle{\bar\psi(0)\bar\psi(t_0)}\rangle =  \int _{-\infty}^{\infty} dt dt' \,f(t')\, f(t-t_0)\, C_0(t-t')
\end{equation}
be the correlation between two measurements separated in time by $t_0$. Here the peaks of the sampling functions of these measurements are 
displaced by $t_0$.
This integral contains a second order pole at $t = t'$, which may be treated by an integration by parts procedure.
Use 
 \begin{equation}
 C_0(t - t')=-\frac{1}{8\pi^2}\frac{\partial^2}{\partial t\partial t'}\log{\left[((t-t')^2\, \alpha^2\right]}
 \end{equation}
to find
\begin{equation}
 C(t_0)=-\frac{1}{8\pi^2}\int dt dt' \,\dot f(t')\, \dot f(t-t_0)\, \log{\left[ (t-t')^2 \,\alpha^2 \right]}\,,
 \end{equation}
which is independent of the value of the arbitrary constant $\alpha$.  This expression is finite, and may be explicitly evaluated for the case of the 
Lorentzian sampling function, Eq.~\eqref{eq:Lorentz} to find
\begin{equation}
  C(t_0) = \frac{(4 - t_0^2)}{4\pi^2\left(t_0^2+4\right)^2}\,.
   \label{eq:C-scalar}
 \end{equation}
 This function is plotted in Fig.~(\ref{fig:C(t0)1}).
Here units in which $\tau =1$ are used, so the time separation $t_0$ is given as a multiple of $\tau$.
Note that $ C(t_0) > 0$ if $t_0 < 2$, so a pair of measurements with a small temporal separation are positively correlated, but larger separations  $t_0 > 2$, 
lead to anti-correlations, $ C(t_0) < 0$. The latter result is expected from the minus sign in Eq.~\eqref{eq:C0}. The mathematical origin of the former result
 is more subtle, but it incorporates the physically reasonable result that two measurements made in rapid succession  are positively correlated.

The probability distribution for operators linear in a free quantum field is a Gaussian
\begin{equation}
 P(\bar\psi) = \frac{1}{\sigma\, \sqrt{2 \pi}}  \exp\left(- \frac{{\bar\psi}^2}{2 \sigma^2} \right) 
 \label{eq:Prob}
 \end{equation}
with variance $\sigma^2 = C(0) = \frac{1}{4\pi^2}$. This means that if we consider a pair of measurements with separation $t_0$, the outcome of the first measurement
is equally likely to be either positive or negative. However, the outcome of the second measurement is more likely to have the same sign as the first if  
$t_0 < 2$, and more likely to have the opposite sign if  $t_0 > 2$. 
A method for implementing this bias in a numerical simulation will be a key topic to be treated in Sect.~\ref{sec:Sims}.

\subsection{The Electromagnetic and Similar Fields}
\label{sec:EM}

The massless scalar field has dimensions of inverse length in units where $\hbar = c =1$. Some other massless fields of physical interest have dimensions of 
inverse length squared,
including the electromagnetic field, and first derivatives of the massless scalar field. In all of these cases, Eq.~\eqref{eq:C0} is replaced by a vacuum correlation function of the form
\begin{equation}
 C_0(t-t') = \frac{\kappa}{4\pi^2(t-t')^4} \,. 
 \label{eq:C-vac}
  \end{equation}
 where $\kappa > 0$ is a constant which depends upon the specific case. Linear field operators will still have a probability distribution of the form 
 of Eq.~\eqref{eq:Prob}.
Note that if we rescale an averaged field by ${\bar\psi} \rightarrow   \sigma {\bar\psi}$ then the rescaled field has variance one. Here it is convenient 
to suppose that this has 
been done, so that we can discuss several cases with different actual variances at once. 

Again we assume time averaging of the field operators with
the Lorenztian function, Eq.~\eqref{eq:Lorentz} with $\tau =1$.
 Now the correlation function with unit variance becomes
\begin{equation}
 K(t_0) = \langle{\bar\psi(0)\bar\psi(t_0)}\rangle =  \frac{1-6t_{0}^{2}+t_{0}^{4}}{(1+t_{0}^{2})^{4}}\,,
 \label{eq:K}
 \end{equation}
which satisfies $K(0) = 1$, as required. 

The function $K(t_0)$ is plotted in Fig.~(\ref{fig:K(t0)}), and in more detail in Fig.~3 of Ref.~\cite{FR2005}. 
It is qualitatively similar in form to $ C(t_0)$ in that it is positive for
$t_0 \alt 0.25$ and negative for $0.25 \alt t_0 \alt 0.8$, representing regions of correlation and anti-correlation, respectively. 

\subsection{Baths of Photons or Gravitons in a Squeezed State}
\label{sec:squeezed}

Sections.~\ref{sec:scalar} and \ref{sec:EM}  have dealt with the vacuum fluctuations of a quantized field, where averaging is essential to define 
finite fluctuations. There is another source of quantum fluctuations arising from particles in a non-classical state, such as a squeezed vacuum. 
Photons or gravitons in such a state can give rise to Brownian motion of test particles. The operator whose fluctuations cause this motion will 
be the electric field for photons and the linearized Riemann tensor for gravitons. In either case, the field fluctuations are described by a correlation
function of the form of Eq.~\eqref{eq:corr}. However, now we are interested in an excited state with a large occupation number and wish to ignore the vacuum
contribution, so we replace the expectation value by the difference between an expectation value in the squeezed state and that in the vacuum.
If only a finite number of modes are excited, this difference is finite when $t=t'$, and time averaging is not needed. In the case that
a single mode of wavenumber  $k$ is occupied, the correlation function with unit variance may be written as
 \begin{equation}
 C_1(t-t') = \cos[k(t-t')]\,,
 \label{eq:sq-corr}
 \end{equation}
an oscillatory function exhibiting alternating correlations and anti-correlations. The Brownian motion of test particles in squeezed states of photons or gravitons
will be discussed further in Ref.~\cite{WHF}

\section{Numerical Simulations with Correlations}
\label{sec:Sims}

There are well-known numerical methods to generate a sequence of otherwise random numbers which obey a specified probability distribution, such as the
Gaussian of Eq.~\eqref{eq:Prob}. An example is the command RandomVariate in the program {\it Mathematica}. However, generating a sequence which 
satisfies a given correlation function is more difficult, but can be done by Monte Carlo methods~\cite{CTY94}, for example. These methods are used in
autoregression analyses in many applications. 

\subsection{Basic method}
\label{sec:Basic}

Here we propose an alternative method for use in simulating quantum fluctuations that will involve shifting the origin of the probability distribution
from which a given outcome will be drawn, in a way depending upon a previous outcome. Suppose that we make a measurement of a field value
for which the probability distribution $P_0(x)$  is of the form of Eq.~\eqref{eq:Prob}. The first outcome, $x_1$, is equally likely to have either sign.
However, if the correlation function is non-zero, the value of  $x_1$ will bias the outcome  $x_2$, of a subsequent measurement, and the magnitude of
 $x_1$ will influence the degree of bias. If the correlation is positive, $C>0$, then  $x_2$ is more likely than not to have the same sign as  $x_1$. 
Similarly, if $C<0$, then  $x_1$ and  $x_2$ are more likely to have opposite signs. 

Our specific proposal is to shift the probability distribution for $x_2$ by an amount proportional to $x_1$:
\begin{equation}
 P(x) \rightarrow P(x -f\, x_1) \, ,
 \label{eq:shift}
 \end{equation}
where $f$ is a constant with $|f| \leq 1$. Thus, $f > 0$ leads to positive correlation, and $f < 0$ to anti-correlation. If the pair of events are separated by
$t_0$ in time with correlation function $C(t_0)$, we need to select $f = f(t_0)$ so that 
 \begin{equation}
 C(f(t_0)) = C(t_0) = \langle x_1\, x_2 \rangle\,, 
 \end{equation}
 the average of the products of pairs of outcomes.
 
 We use the following trial form for $C(f)$:
  \begin{equation}
 C(f) = a\, \tan(b \, f)\,,
 \end{equation}
 where $a$ and $b$ are constants. This functional form has the following properties: 1) $C(0) = 0$, so $f=0$ describes lack of correlation;
 2) The range of $C$ is infinite, while that of $f$ is finite;  $-\infty < C < \infty$  and $ -1 < 2\, b \,f/\pi < 1$; 
 3) It is an odd function of $f$, which allows the possibility to treat negative correlations as well as positive ones.
 For each of the three correlation functions
 discussed in Sect.~\ref{sec:corr}, we have found numerical values of $a$ and $b$ which yield reasonable simulated correlation functions.

 \subsubsection{Massless Scalar Field}
\label{sec:scalar-fit}

\begin{figure}[hbt!]
    \centering
    \includegraphics[width=8cm]{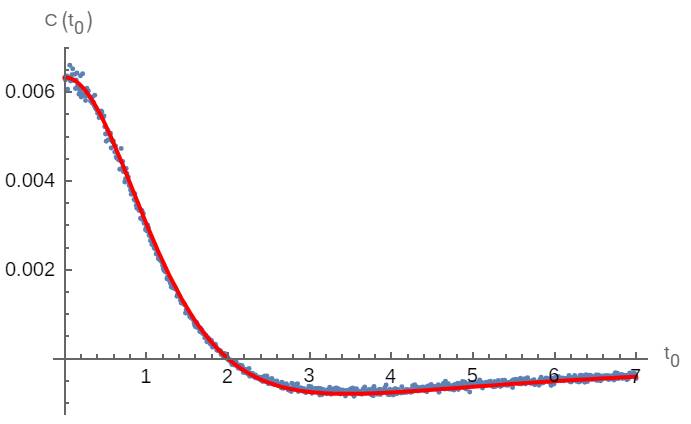}
    \caption{The correlation function $C(t_0)$ given by  Eq.~\eqref{eq:C-scalar}  (red) is plotted against a simulated function  (blue)}. 
    \label{fig:C(t0)1}
\end{figure}

Here the correlation function, $C(t_0)$ is given by Eq.~\eqref{eq:C-scalar}, and we use
 \begin{equation}
 f(t_{0}) = \frac{1}{b} \arctan\left(\frac{4 - t_{0}^{2}}{4\pi\, a\,(4 + t_{0}^{2})^{2}} \right)\,.
 \end{equation}
with
 \begin{equation}
 a \approx 0.01404 \qquad b \approx 1.58\,.
 \end{equation}
 The analytic and simulated forms of this correlation function are plotted together in Fig.~\ref{fig:C(t0)1}. In this and the following plots, 
 801 points are plotted and 20,000 steps are used to create each point.
 
\subsubsection{The Electromagnetic and Similar Fields}
\label{sec:EM-fit}

Here the correlation function, $K(t_0)$ is given by Eq.~\eqref{eq:K}, and we use
 \begin{equation}
 f(t_{0}) = \frac{1}{b} \arctan\left(\frac{1-6t_{0}^{2}+t_{0}^{4}}{a(1+t_{0}^{2})^{4}}\right)
 \end{equation}
with
 \begin{equation}
  a \approx 0.672 \qquad b \approx 1.59\,.
 \end{equation}

\begin{figure}[hbt!]
    \centering
    \includegraphics[width=8cm]{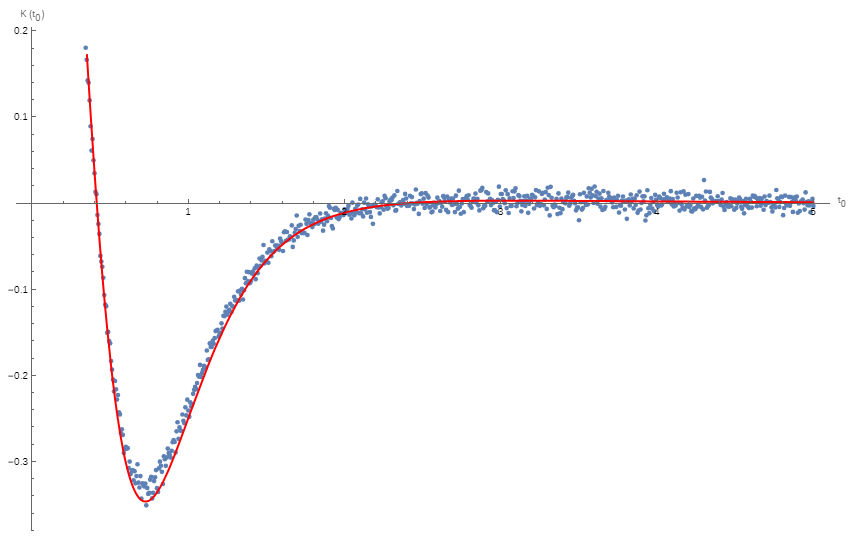}
    \caption{The correlation function $K(t_{0})$ (red) and its simulated form (blue) are plotted.}
    \label{fig:K(t0)}
\end{figure}

 The analytic and simulated forms of this correlation function are plotted together in Fig.~\ref{fig:K(t0)}.

\subsubsection{Baths of Photons or Gravitons in a Squeezed State}
\label{sec:squeezed-fit}

\begin{figure}[hbt!]
    \centering
    \includegraphics[width=8cm]{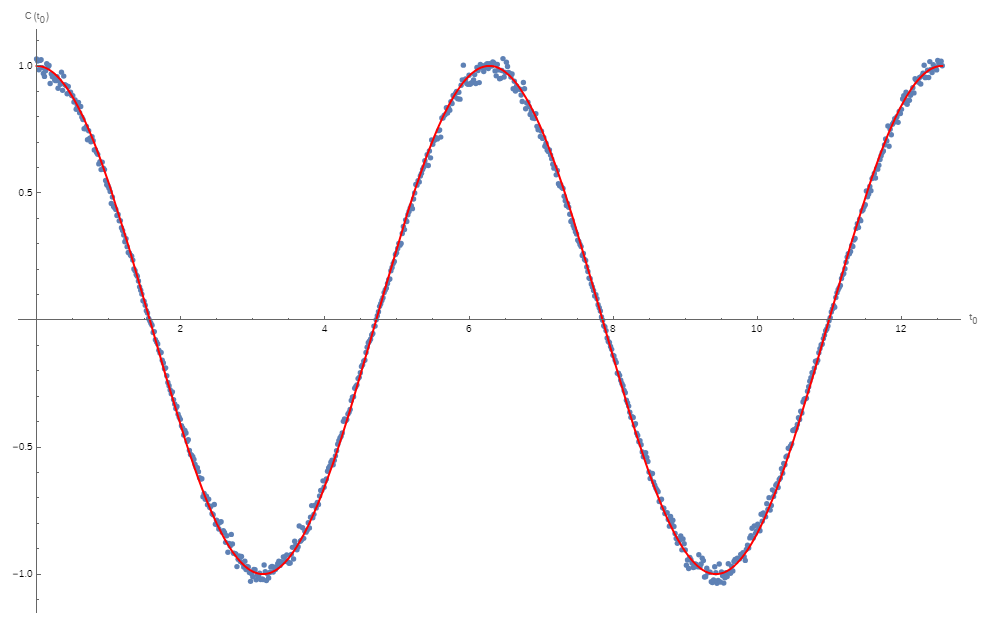}
    \caption{The correlation function $C_1(t_{0})$ (red) and its simulated form (blue) are plotted.}
    \label{fig:cos}
\end{figure}

Here the correlation function, $C_1(t_0)$ is given by Eq.~\eqref{eq:sq-corr}, and we use
 \begin{equation}
 f(t_{0}) = \frac{1}{b} \arctan\left(\frac{\cos(k \, t_{0})}{a}\right)
 \end{equation}
with
  \begin{equation}
  a \approx 0.672 \qquad b \approx 1.59\,.
 \end{equation}
   The analytic and simulated forms of this correlation function are plotted together in Fig.~\ref{fig:cos}.
   
   In all three cases, we see that the simulations and analytic forms of the correlation functions agree well. We have also used a modified version of 
our procedure in which we start with the derivative $C'(t_0)$ of the analytically derived correlation function, obtain a numerical simulation of this derivative, 
and then numerically integrate the result to find simulated functions which agree reasonably well with the simulations plotted in Figs.~\ref{fig:C(t0)1}, 
~\ref{fig:K(t0)}, and \ref{fig:cos}.

 \section{Application to the Brownian Motion of Test Particles}
\label{sec:Brownian} 
   
One possible application of the method developed in Sect.~\ref{sec:corr} could be to the study  of the effects of field fluctuations on Brownian motion,
the fluctuations in test particle trajectories.  This topic was discussed using analytic correlation functions in many papers, including Refs~\cite{Yf04,PF11}.
Quantum electric field fluctuations create a fluctuating force on a test charge and hence cause the linear momentum of the charge to fluctuate. In the
absence of an external source of energy, the mean squared momentum cannot on average grow in time. In this case, energy   conservation is
enforced by the anti-correlations of the electric field fluctuations. The particle may temporarily acquire energy from a fluctuation, but soon an 
anti-correlated fluctuation will tend to take this energy away. This feature is encoded in the correlation function by the vanishing of its integral over all time.
For the case of the quantized electric field, 
 \begin{equation}
  \int _{0}^{\infty} K(t_0)\, dt_0  = 0\,.
 \end{equation}
   
   Numerical simulations may give insight into the behavior of test particles in baths of squeezed photons or squeezed gravitons, and will be a topic of
   future research. For the purpose of this paper, we consider a simple example of a test particle coupled to the fluctuating massless scalar field. At each 
   step the value of the field $\bar\psi$  found determines the sign and magnitude of the particle's displacement for that step. If the field fluctuations
   were uncorrelated, this would constitute a random walk in which the mean squared displacement grows proportionally with the square root of the 
   number of steps. Correlations are expected to either enhance or retard this rate of growth, depending upon the sign of the correlation. 
   We have performed simulations for the cases  $f= -0.5$, $f=0$ ,and $f=0.5$, and found square root growth in all cases for step number $N$ in the range
   $0< N <100$. specifically
    \begin{equation}
 y \approx \sqrt{ 0.68 N +0.039} \qquad  f= -0.5 \,,
 \end{equation}
  \begin{equation}
 y \approx \sqrt{ 1.32 N +0.092} \qquad  f= 0 \,,
 \end{equation}
 and
  \begin{equation}
 y \approx \sqrt{ 5.35 N +8.13} \qquad  f= 0.5 \,.
 \end{equation}
We see that the uncorrelated case, $f=0$, grows faster than that of anti-correlation,
   $f= -0.5$, but slower than the case of positive correlation, $f= 0.5$. The fact that we still find some growth in the case of anti-correlation is probably 
   due to the limited nature of this simulation. Here we are assuming that  each step is correlated only with the preceding step, and not with earlier steps.
   A more realistic multi-step simulation is a topic for future research.

\section{Summary}
\label{sec:final}

In this paper we have proposed a method to numerically simulate quantum field fluctuations with a given temporal correlation. This potentially allows
simulations of the subtle effects of correlation and anti-correlation which quantum fields exhibit. Here our attention was restricted to linear fields with
a Gaussian probability distribution, and our method involves a displacement of the Gaussian which depends upon the 
outcome of  a previous measurement. We were able to use this method to numerically implement correlations in several explicit cases. We also used
this method for a simplified treatment of the effects of correlations on a random walk. That treatment involved only one-step correlations between an 
event and the immediately prior event. We hope to generalize our treatment to include multi-step correlations with several prior events.

We also plan to extend our method to cases of non-Gaussian probability distribution. This will be of interest in simulating quantum stress tensor
fluctuations, which can have non-Gaussian probability distribution which falls more slowly than exponentially, and depends upon the details of the 
measurement sampling function $f(t)$~\cite{FFR10,FF15}.

\begin{acknowledgments} 
This work was supported in part  by the National Science Foundation under Grant PHY-2114024.
\end{acknowledgments}
.\\
$^\dagger$\href{mailto:emily.taylor@tufts.edu}{emily.taylor@tufts.edu}\\
$^\ddagger$\href{mailto:ford@cosmos.phy.tufts.edu}{ford@cosmos.phy.tufts.edu}\\
$^*$\href{mailto:yenchs@rpi.edu}{yenchs@rpi.edu}\\

\end{document}